# Mechanistic home range models and resource selection analysis: a reconciliation and unification

Paul R. Moorcroft[1]

&

Alex Barnett[2]

[1] OEB Dept, Harvard University, 22 Divinity Ave, Cambridge MA 02138
Phone: (617)-496-6744 Fax: 617-495-9484 Email: paul_moorcroft@harvard.edu

[2] Department of Mathematics, 6188 Kemeny Hall, Dartmouth College, Hanover, NH, 03755

## ABSTRACT

**In the three decades since its introduction, resource selection analysis (RSA) has become a widespread method for analyzing spatial patterns of animal relocations obtained from telemetry studies. Recently, mechanistic home range models have been proposed as an alternative framework for studying patterns of animal space use. In contrast to RSA models, mechanistic home range models are derived from underlying mechanistic descriptions of individual movement behavior and yield spatially-explicit predictions for patterns of animal space-use. In addition, their mechanistic underpinning means that, unlike RSA, mechanistic home range models can also be used to predict changes in space-use following perturbation. In this paper, we develop a formal reconciliation between these two methods of home range analysis, showing how differences in the habitat preferences of individuals give rise to spatially-explicit patterns of space-use. The resulting unified framework combines the simplicity of resource selection analysis with the spatially-explicit and predictive capabilities of mechanistic home range models.**

## INTRODUCTION

Since its introduction in the 1980s, resource selection analysis (RSA) has become a widespread method for identifying underlying environmental correlates of animal space-use patterns (Manly 1974, Johnson 1980). In contrast to earlier descriptive methods of home range analysis, such as the minimum convex polygon, bivariate normal and kernel methods that simply summarize observed spatial patterns of relocations (see MacDonald 1980a, Worton 1987, and Kernohan et al. 2001 for reviews), conventional RSA uses a spatially-implicit frequentist approach to identify habitats that are used disproportionately in relation to their occurrence.

Ratios of habitat utilization relative to habitat availability provide a simple estimate of habitat selection (Figure 1). More commonly however, resource selection models are specified in terms of the probability $P_j$ of obtaining a relocation in a given habitat type $j$:

$$P_j = \frac{A_j w_j}{\sum_{k=1}^{m} A_k w_k} \qquad (1)$$

where $w_j$ is the selective value of habitat $j$ relative to other habitats ($j=1..m$, where $m$ is the number of habitat types) and $A_j$ is the availability of habitat $j$ on the landscape. The collection of resource selection values for all habitats within a landscape $\{w_j\}$ ($j=1..m$) is known as the resource selection function (Manly et al. 1993). Eq. (1) is generally preferred over simpler ratio-based estimates of habitat selectivity because of its improved statistical properties, having smaller variance and being less subject to bias (Arthur et al. 1996).

As results from numerous studies have shown, RSA can be successfully used to identify associations between animal space-use and habitat types as well and other forms of environmental heterogeneity, such as topography and resource availability, yielding insight into the underlying causes of animal space use (see Manly et al. 1993; Boyce & MacDonald 1999; Cooper & Millspaugh 2001; Erickson et al. 2001 for reviews).

More recently, mechanistic home range models have been proposed as an alternative framework for analyzing patterns of animal home ranges (Moorcroft & Lewis 2006, Moorcroft et al. 2006). In contrast to the spatially-implicit, frequentist nature of RSA models such as Eq. (1), mechanistic home range models develop spatially-explicit predictions for patterns of animal space-use, by, in the words of Millspaugh and Marzluff (2001), "modeling the movement process". Mathematically, this involves characterizing the fine-scale movement behavior of individuals as a stochastic movement process that is defined in terms of a redistribution kernel $k(\mathbf{x},\mathbf{x}',\tau,t)$, where $k(\mathbf{x},\mathbf{x}',\tau,t)d\mathbf{x}$ specifies the probability of an animal located at $\mathbf{x}'$ at time $t$ moving to a location between location $\mathbf{x}$ and $\mathbf{x}+d\mathbf{x}$ in the time interval $\tau$.

Relevant behavioral and ecological factors influencing the movement of individuals can be incorporated into the redistribution kernel that defines the fine-scale stochastic movement process. For example, in a recent analysis of coyote home ranges in Yellowstone, Moorcroft et al. (2006) developed a "prey availability plus conspecific avoidance" (PA+CA) mechanistic home range model in which individuals exhibit: (i) an avoidance response to encounters with foreign scent marks (Figure 2a,b), (ii) an over-marking response to encounters with foreign scent marks, and (iii) a foraging response to prey availability in which individuals decreased their mean step length in response to small mammal abundance (Figure 2a,c).

From this description of fine-scale movement behavior, it is then possible to derive probability density functions for the expected spatial pattern of home ranges that result from individuals moving on a landscape according to the underlying rules of movement. For example, in the case of the mechanistic home range model used by Moorcroft et al. to analyze coyote movements, the stochastic foraging response and responses to scent-marks yield the following equations for the expected steady-state pattern of space use:

$$\frac{\partial u^{(i)}(\mathbf{x},t)}{\partial t} = \underbrace{-\nabla\cdot\left[e^{-\alpha h(\mathbf{x})}\nabla u^{(i)}\right]}_{\text{random motion}} + \underbrace{\nabla\cdot\left[e^{-\alpha h(\mathbf{x})}\beta(\mathbf{x})\vec{\mathbf{x}}_i u^{(i)}\sum_{j\neq i}^{n_{pack}} p^{(j)}\right]}_{\text{scent-mark avoidance}} \underbrace{-\nabla\cdot\left[e^{-\alpha h(\mathbf{x})}u^{(i)}\nabla h\right]}_{\text{directed movement towards areas of high resource density}} = 0 \quad , \tag{2}$$

where $\nabla = \left(\frac{\partial}{\partial x},\frac{\partial}{\partial y}\right)$,

and

$$\frac{dp^{(i)}(\mathbf{x},t)}{dt} = \underbrace{u^{(i)}\left[1+m\sum_{j\neq i}^{n_{pack}} p^{(j)}\right]}_{\text{scent-mark deposition}} - \underbrace{p^{(i)}}_{\text{scent-mark decay}} \quad , \tag{3}$$

where $u^{(i)}(\mathbf{x},t)$ is the expected space-use of individuals in pack $i$, $p^{(i)}(\mathbf{x},t)$ is the scent-mark density of individuals in pack $i$, and $h(\mathbf{x})$ is the spatial distribution of prey availability. The coefficients $\alpha$ and $\beta$ reflect the underlying characteristics of individual movement behavior. Specifically, $\alpha=2\rho_1$, where $\rho_1$ is the mean step length of the individual, and $\beta$ reflects the strength of the avoidance response to foreign scent marks (Figure 2b,c). Figure 2d shows the fit of the PA+CA space-use equations to the observed spatial distribution of relocations of five adjacent coyote packs in Yellowstone National Park home ranges. As the figure illustrates, the model captures the influences of resource availability and the presence of neighboring groups on the patterns of space use within the region.

Mechanistic home range models address several limitations of RSA. First, as the fit of the PA+CA model shown in Figure 2d illustrates, in addition to incorporating the effects of underlying landscape heterogeneities such as prey density, mechanistic home range models can also incorporate the influence of conspecifics that can significantly influence patterns of animal space use. Second, a critical step in any RSA is defining the region within a landscape that constitutes available habitat. The spatially-implicit, frequentist nature of conventional RSA models such as Eq. (1) means that all areas within the pre-defined region defined as “available habitat” are assumed to be equally accessible to individuals. On real landscapes, the patchy spatial distribution of habitats and resources such as that seen in Figure 2d, means that individuals are frequently required to traverse less favorable habitats in order to move between preferred areas. In conventional RSA, the time individuals spend traversing unfavorable habitat registers as a degree of selection for that habitat type rather than as a constraint imposed by the spatial distribution of habitats on the landscape. In contrast, the mechanistic and spatially explicit nature of mechanistic home range models avoids the need to define available habitat, *a priori*, since the underlying model of individual movement behavior (Figure 2) determines the likelihood and feasibility of an individual moving to a particular location given its current position. Thus, by explicitly modeling the process of individual-level movement, mechanistic home range models naturally capture the influence of spatial constraints on patterns of space use by individuals. Finally, their mechanistic nature means that mechanistic home range models can be used to predict patterns of space-use following perturbation. For example, Moorcroft et al. (2006) showed that their PA+CA model correctly captured the shifts in patterns of space-use that occurred following the loss of one of the packs in the study region. For further discussion of these issues, see Moorcroft et al. (2006) and Moorcroft & Lewis (2006).

The above considerations imply that resource selection analysis (RSA) and mechanistic home range models constitute two distinct frameworks for analyzing patterns of animal space use. However, as we show below, some recent developments now permit a formal reconciliation and unification of these two seemingly disparate methods for analyzing patterns of animal space-use.

The first important development toward reconciling mechanistic home range models and RSA came in an analysis of polar bear relocations by Arthur et al. (1996), who argued that rather than assuming a fixed measure of habitat availability across the entire study region, a more appropriate measure was the availability habitats within a circle centered on the individual's current location, and whose radius $R$ corresponded to the maximum distance the individual was likely to travel in the time between successive relocations. They incorporated this into an RSA model by using modified version of Eq. (1), in which habitat availability varies between relocations:

$$P_{ij} = \frac{A_{ij} w_j}{\sum_{k=1}^{m} A_{ik} w_k}, \qquad (4)$$

where $P_{ij}$ is the probability of choosing habitat $j$ for the $i^{\text{th}}$ move, $w_j$ is the habitat selection parameter for habitat *j*, and $A_{ij}$ is the proportional availability of habitat *j* associated with relocation *i*, calculated as the fraction of the area within distance *R* of location *i* that is of type *j*.

The second development came in a paper by Rhodes et al. (2005) who proposed an extension of Arthur et al.'s (1996) approach to defining available habitat. Eq. (4) incorporates spatial variation in habitat availability; however, like Eq. (1), it is written in terms of the probability of observing a relocation of a given habitat type. Rhodes et al. (2006) re-cast Arthur et al.'s model Eq. (4) in terms of the probability of an individual moving from location given location *a* to a subsequent location *b*:

$$P(a \to b) = \frac{\phi(a,b)\sum_{j=1}^{m} w_j I(b,j)}{\sum_{j=1}^{m} w_j \int_{I(c,j)=1} \phi(a,c)\, dc} \qquad (5)$$

where *I(b,j)* is an indicator function that takes the value 1 when location *b* is of type *j* and zero otherwise, and $\phi(a,b)$ is given by:

$$\phi(a,b) = \begin{cases} \dfrac{1}{\pi R^2} & \text{if} \quad r_{ab} \leq R \\ 0 & \text{otherwise} \end{cases} \qquad (6)$$

where $r_{ab}$ is the distance between locations *a* and *b*, and *R* is to the maximum distance an individual is likely to travel between successive relocations, which defines the area of available habitat at location *a*.

Rhodes et al.'s motivation for casting Arthur et al.'s model into the form of Eq. (5) was twofold. First, they argued for a different functional form for $\phi(a,b)$ than Eq. (6), namely an exponential distribution $\phi(a,b) = \lambda \exp(-\lambda r_{ab})/2\pi r_{ab}$. Second, they introduced distance from the center of the individual's home range as a spatial covariate of the resource selection parameter, that is $w_j$ becomes a function $w_j(x)$). Rhodes et al. termed this a "spatially-explicit habitat selection model" to reflect the spatially-varying resource selection coefficient. Note however that, unlike mechanistic home range models, resource selection models such as Eq. (5) do not give rise to predictions for actual patterns of space-use by individuals.

Switching from a model defined in terms of the probability of observing a relocation in a given habitat type to a model defined in terms of the probability of an individual moving between its current location and its subsequent location has, however, a third important consequence: Eq. (5), unlike Eq. (4), constitutes a redistribution kernel for the fine-scale movement behavior of individuals. In other words, RSA models of the form of Eq. (5) are, in effect, "modeling the movement process". The above observation implies that it should be possible to establish a formal connection between RSA models of the form of Eq. (5) and corresponding mechanistic home range model. Below we consider a simple pedagogical example that demonstrates this is indeed the case.

## ANALYSIS

Consider an individual living on a one-dimensional landscape whose relative preference for different habitats can be expressed by a resource selection function $w(x)$ (Figure 3a). Suppose further that in the absence of habitat preference (i.e. $w(x)$ constant), the individual moves to the right or left of its current position during time interval $\tau$, with a distribution of displacements $\phi_\tau(q)$ where $q = x - x'$ is the displacement between the individual's current location *x'* and its subsequent location *x* (Figure 3b)[1]. Since $\phi_\tau(q)$ is a probability density function $\int_{-\infty}^{\infty} \phi_\tau(q)dq = 1$.

The probability density of the individual moving to location *x* from its initial location *x'*, during time interval $\tau$, in this landscape with varying preference is then given by our model redistribution kernel

$$P(x' \to x) = k_\tau(x,x') = \frac{\phi(x-x')w(x)}{\int_{-\infty}^{\infty} \phi(x''-x')w(x'')dx''}. \qquad (7)$$

Note that Eq. (7) has the same form as Eq. (5), with $w(x) = \sum_{j=1}^{m} w_j I(x,j)$ Note that the probability of moving from *x'* to *x* in the absence of habitat preference is determined only by the difference between *x* and *x'* and that the preference function *w* is evaluated at the location to which the individual moves, rather than its current location. In this example we assume that an individual's redistribution kernel does not vary in time, and thus the dependency on time *t* can be dropped.

Defining *u(x,t)dx* as the probability that the individual is located between *x* and *x+dx* at time *t,* we can write an equation that summarizes all the possible ways that an individual located at $x'$ can arrive within the interval *[x, x+dx)* at time $t+\tau$

$$u(x,t+\tau) = \int_{-\infty}^{\infty} k_\tau(x,x')\, u(x',t)dx' \qquad (8)$$

Eq. (8) is converted into an equation for the expected pattern of space use by the individual by expanding the right-hand side using a Taylor series and then considering

[1]. Note that in contrast to Moorcroft et al., where the fine-scale movement behavior in two dimensions is described in terms of a sequence of movements of length $\rho_\iota$ and direction $\phi_\iota$ *(i=1...m)* (see Figure 2), here we describe the fine-scale movement behavior of an individual moving in a single space dimension in terms of a sequence of displacements $q_\iota$ *(i=1...m)* that have both a magnitude and a sign.

the limit as $\tau \to 0$, yielding the following advection-diffusion equation:

$$\frac{\partial u(x,t)}{\partial t} = -\frac{\partial}{\partial x}\left[c(x)u(x,t)\right] + \frac{\partial^2}{\partial x^2}[d(x)u(x,t)], \tag{9}$$

where the advection and diffusion coefficients, *c(x)* and *d(x)* respectively, are given by

$$\begin{aligned} c(x) &= \lim_{\tau\to 0}\frac{1}{\tau}\int_{-\infty}^{\infty}(x-x')\,k_\tau(x,x')dx' \quad \text{and} \\ d(x) &= \lim_{\tau\to 0}\frac{1}{2\tau}\int_{-\infty}^{\infty}(x-x')^2\,k_\tau(x,x')dx'. \end{aligned} \tag{10a,b}$$

Details of the derivation can be found in the Appendix. Since *u(x,t)* is a probability density function, the normalization

$$\int_\Omega u(x,t)\,dx = 1, \tag{11}$$

where Ω is the region over which the individual is able to move, is preserved for all future times *t*.

Inserting Eq. (7) into Eq.s (10a,b) (see Appendix) yields the following equations for the coefficients *d* and *c*:

$$\begin{aligned} c(x) &= \lim_{\tau\to 0}\frac{M_2(\tau)}{\tau}\frac{w_x(x)}{w(x)}, \\ d(x) &= \lim_{\tau\to 0}\frac{M_2(\tau)}{2\tau}, \end{aligned} \tag{12a,b}$$

where the second moment is $M_2(\tau) = \int \rho^2\,\phi_\tau(x)dx$, and $w_x = \frac{dw}{dx}$.

Thus we see that a simple spatially-explicit resource-selection model yields an advection-diffusion equation Eq. (9) for the expected location of an individual. Note that while the advection term (Eq. 12a) varies in space, the magnitude of the diffusion coefficient (Eq. 12b) is constant.

Inserting Eq.s (12a,b) into Eq. (9) we can derive an approximation for the expected steady-state pattern of space use $u^*(x) = \lim_{t\to\infty} u(x,t)$. The result for the case of a smooth continuous preference function *w(x)* is

$$u^*(x) = \frac{1}{W_0} w(x)^2, \qquad (13)$$

with the normalization constant $W_0 = \int_\Omega w(x)^2 dx$.

The details of the derivation are given in the Appendix. In other words, the steady-state pattern of space-use by an individual is given by the normalized square of its resource selection function *w(x)*.

Figure 4a shows a plot of Eq. (13) and a numerical solution of Eq.s (7) and (8) for the case of an individual who in absence of habitat preference moves with exponential distribution of step lengths and with an equal probability of moving in either direction[2]. As can be seen in the figure, Eq. (13) captures the pattern of space-use arising from the underlying habitat preferences of the individual.

Eq. (13) is an approximation that technically holds only when variation in *w(x)* is at spatial scales large relative to the characteristic width of the individual's distribution of displacements $\phi_\tau(q)$. Figure 4b shows a case where the individual's resource selection function is discontinuous. In this case, Eq. (13) does not accurately capture the pattern of space-use in the region of the discontinuity; however, the errors are localized, and thus Eq. (13) still reasonably describes the overall pattern of space use.

[2] Mathematically, this is equivalent to an individual with a Laplace distribution of displacement distances *q*.

## DISCUSSION

The analysis presented here demonstrates that resource selection models of the form of Eq. (5) proposed by Rhodes et al. (2005) constitute an underlying stochastic movement process, and thus can be used to formulate corresponding mechanistic home range models which predict the expected patterns of space-use that result from the underlying habitat preferences. As we would expect, increasing preference for a given habitat type (higher $w(x)$) gives rise to increasing space-use in the preferred habitats relative to the less preferred habitats (Figure 4).

What is surprising however, is that the relative intensity of space-use by an individual at a given location is governed by the square of its preference function for that location. While Eq. (13) is an approximation that technically only holds when the preference function varies on spatial scales larger than the individual's redistribution kernel as seen in Figure 4b it still captures the overall pattern space of space use even when an individual's resource selection function changes rapidly in space. The mathematical explanation for this non-intuitive result is that the equilibrium pattern of space use $u^*(x)$ is governed by the relative strength of the advection term relative to magnitude of the diffusion term, the squared term arising because the factor of 2 in the denominator of the diffusion coefficient Eq. (12b) (see Appendix for further details).

A more biological explanation is as follows. When viewed in terms of a model for individual movement behavior, resource selection models of the form of Eq. (5) reflect a specific assumption about how environmental factors such as habitat type influence an animal's movement: the influence of habitat preference occurs by generating a differential probability of an individual moving in a given movement direction. It is the bias in this probability that gives rise to the positive advection term (Eq. 12a), whose sign and magnitude is governed by the relative gradient in the animal's resource selection function, with preferential movement towards preferred habitat types. It is this preferential movement up gradients of habitat preference relative to the random component of motion (represented by the diffusion term) that determines the individual's equilibrium pattern of space use $u^{*}(x)$.

In mechanistic home range models, environmental and biological factors affecting an individual's movement behavior can influence either, or both, its distribution of movement directions, and its distribution of movement distances. For example, in the PA+CA model used by Moorcroft et al. to analyze coyote home range patterns in Yellowstone (see Introduction), encounters with foreign scent marks alter an individual's distribution of movement directions (Figure 2b). In contrast, consistent with observations of coyote foraging responses (Laundre and Keller 1981), prey availability does not influence an individual's distribution of movement directions, but instead influences its distribution of movement distances, with the mean step length of individuals declining as an exponential function of the prey availability encountered (Figure 2c). These two qualitatively different forms of movement response give rise to different terms in the equations for expected patterns of space use: conspecific avoidance response gives rise to an advective term, indicating directed movement towards its home range center, while the foraging response gives rise to a spatially varying diffusion term (see Eq. 2).

The above discussion implies that RSA models such as Eq. (5) embody one aspect of how animals can respond to factors affecting their movement, namely, by changing their distribution of movement directions, and giving rise to directed movement terms in the space-use equations of the corresponding mechanistic home range model. Changes in an animal's distribution of movement distances such as the foraging response in the PA+CA model of Moorcroft et al. (2006) constitute a second mechanism by which animals respond to their environment -- one that is not represented in current resource selection models. As work by a number of authors, including Okubo (1980), Kareiva and Odell (1982), Turchin (1991, 1998) has shown, the magnitude of this second form of movement response is governed by spatial variation in the mean-squared displacement of individuals that determines spatial variation in the diffusion coefficient in space-use equations such as Eq. (4). Together, these two qualitatively different forms of movement response – preferential movement in particular directions, and spatial variation in mean-squared displacement -- determine the relative intensity of space-use in different areas.

Finally, while the example considered here of an individual moving on a one-dimensional landscape is clearly idealized, previous analyses (Moorcroft and Lewis 2006) suggest that our reconciliation between a resource selection model and a corresponding mechanistic home range model also applies to the more biologically-relevant case of individuals moving on two-dimensional landscapes comprised of multiple habitat types. In the long term, this ability to translate resource selection-based analyses of patterns of animal relocations into corresponding mechanistic home range models offers the promise of a unified framework for analyzing patterns of animal space use.

## ACKNOWLEGMENTS

The authors thank Mark Lewis for his suggestions and comments on this manuscript.

(a)

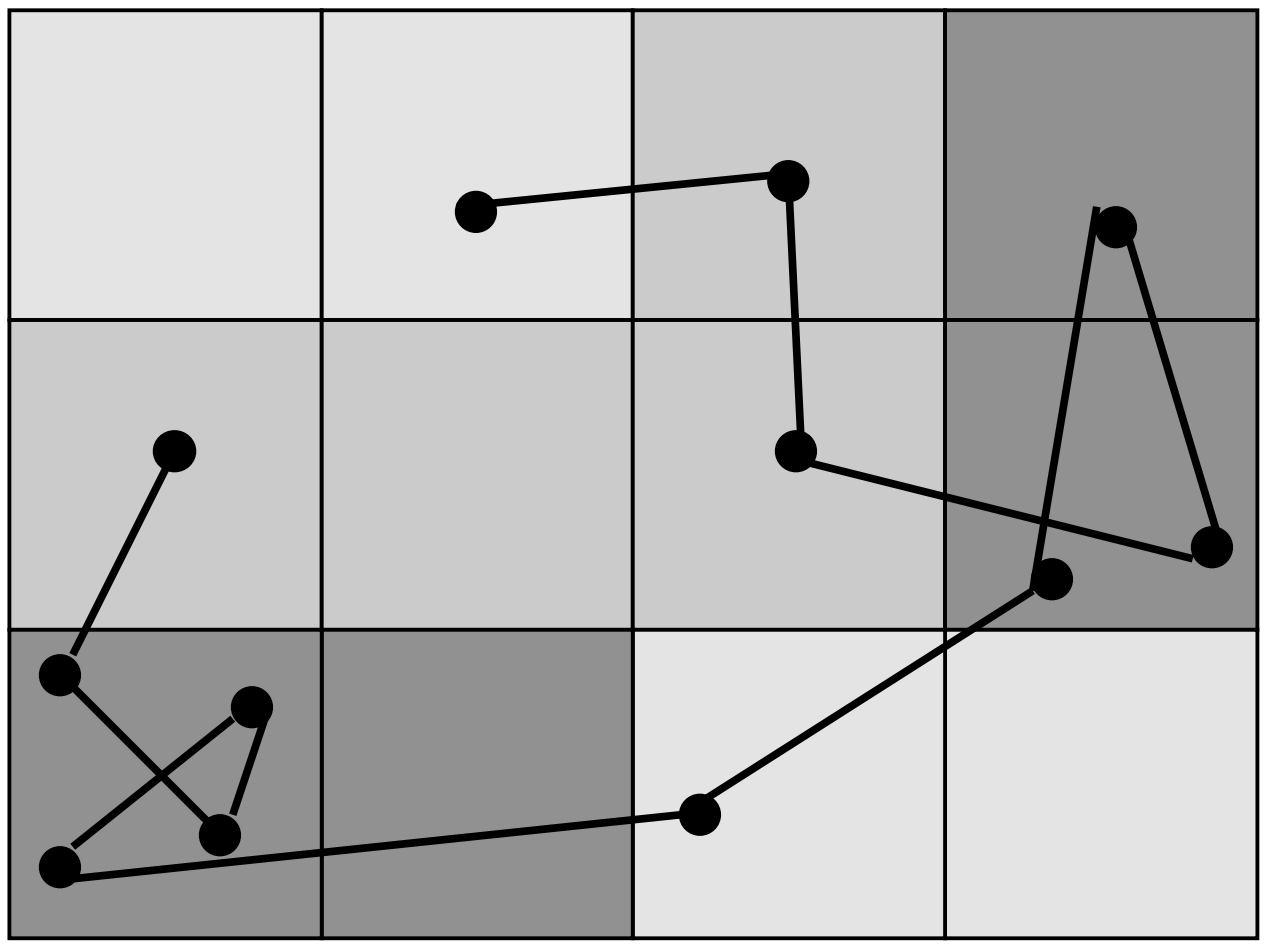

(b)

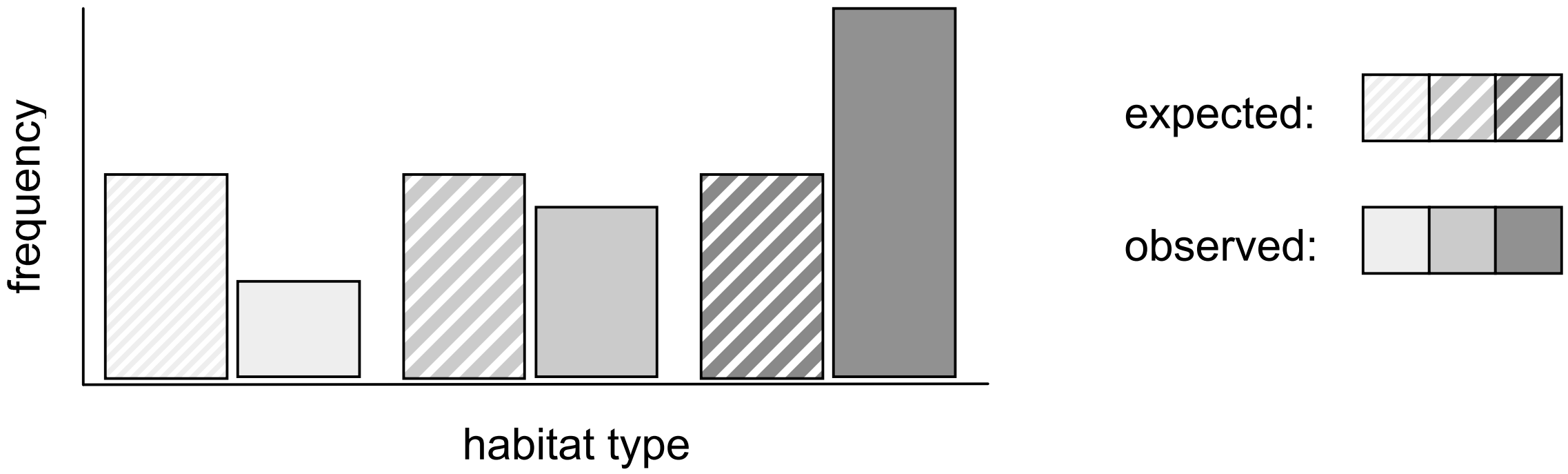

Figure 1. Schematic illustrating the resource selection analysis (RSA) approach to analyzing patterns of animal space-use. (a) Shaded squares represent an idealized landscape comprised of three equally abundant habitat types. Black lines represent the movement trajectory of an individual as it traverses the landscape with points representing fixed-interval relocations of the individual. Since the three habitat types that comprise the landscape plotted in (a) are equally abundant, in the absence of preference, equal numbers of relocations would be expected to be obtained in each habitat, as indicated by the hatched bars in panel (b). The actual distribution of relocations, indicated by the solid bars in panel (b), shows that the individual exhibits a preference for the dark grey habitat type.

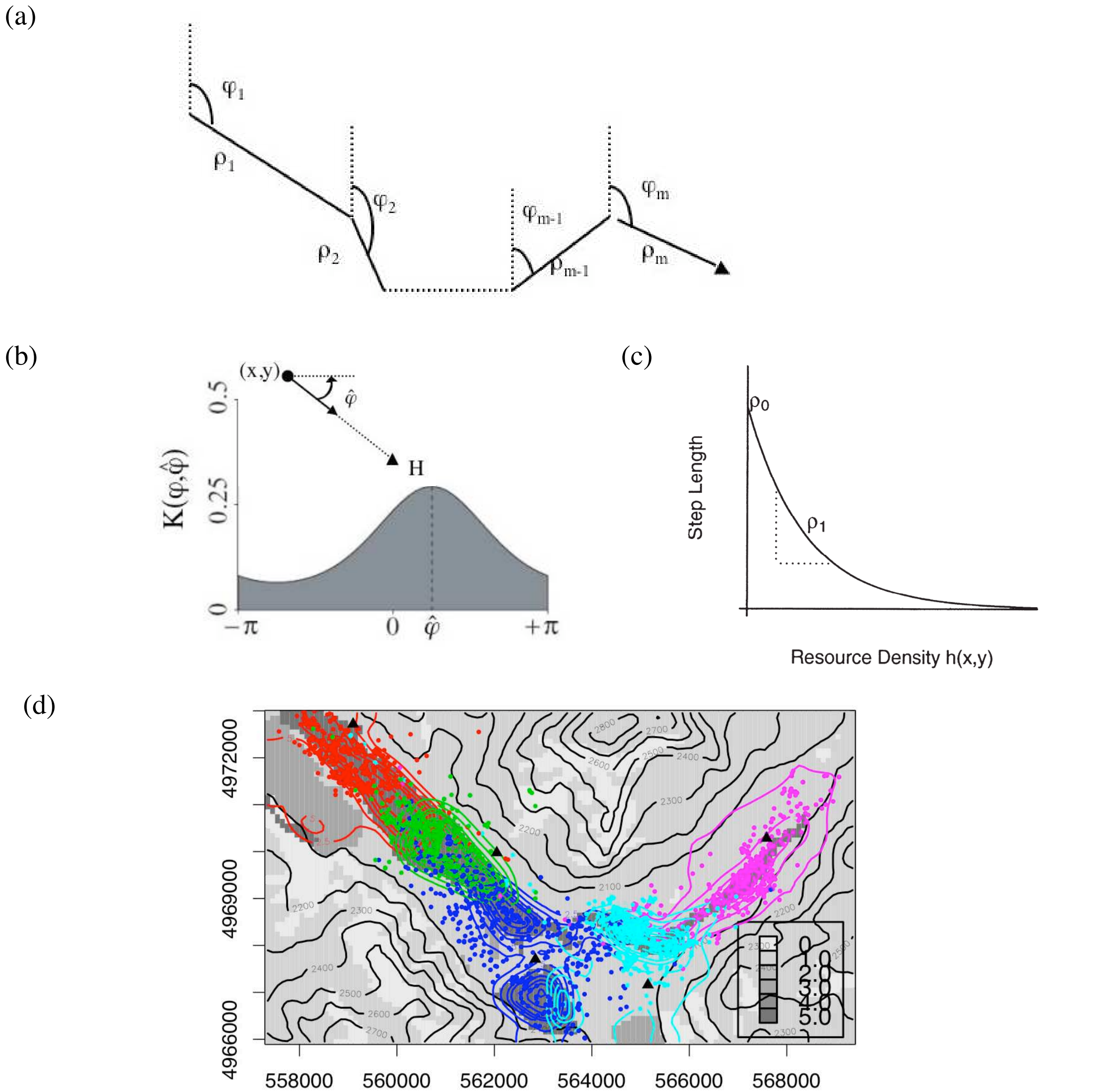

Figure 2. (a) Schematic illustrating the model of individual movement behavior underlying mechanistic home range models. The movement of trajectory of individuals is characterized as a stochastic movement process, defined in terms of sequences of movements *(i=1..m)* of distance $\rho_i$ and directions $\varphi_i$ drawn from statistical distributions of these quantities that are influenced by relevant factors affecting the movement behavior of individuals. (b) and (c): movement responses incorporated in the PA+CA mechanistic home range model used by Moorcroft et al. (2006) to characterize coyote home range patterns in Yellowstone. (b) Individuals display a conspecific avoidance response, in which encounters with foreign scent marks give rise to a non-uniform distribution of movement directions K(φ), with individuals preferentially moving towards the center of their home range (H) whose direction from their current position is indicated by the vector $\hat{\varphi}$. The magnitude of the avoidance response is governed by the density of foreign scent marks encountered and the value of an avoidance parameter $\beta$. (c) Consistent with observations of carnivore movement behavior, increasing prey availability causes the mean step length of individuals to decrease. $\rho_0$ is the mean distance between successive relocations in the absence of resources and $\rho_1$ governs the rate at which the mean step length of individuals decreases with increasing resource density $h(x,y)$. (d) Colored contour lines showing fit of the prey availability plus conspecific avoidance (PA+CA) home range model (Equations 2 and 3) to relocations (filled circles) obtained from 5 adjacent coyote packs in Lamar Valley Yellowstone National Park. The home range centers for each of the packs are also shown (▲), and the grayscale background indicates small mammal prey density (kg ha$^{-1}$) in the different habitat types.

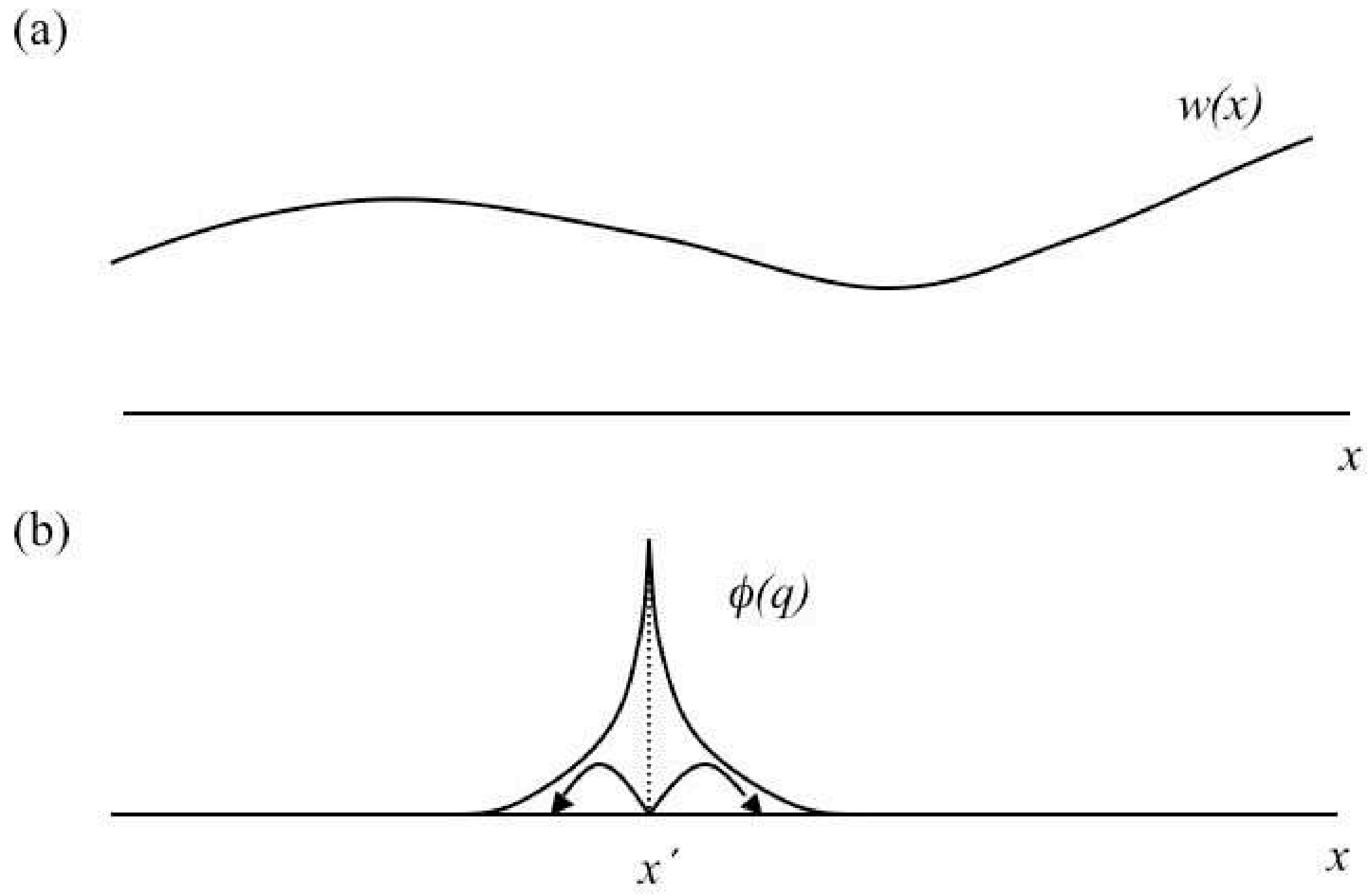

Figure 3 Schematic illustrating a simple resource-selection based mechanistic home range model for an animal moving in a single space dimension. Panel (a) shows the spatially dependent resource selection function *w(x)*. Panel (b) shows the individual's distribution of movement distances *φ(q)*, where *q* is the displacement between the individual's current location *x'* and its subsequent location *x*.

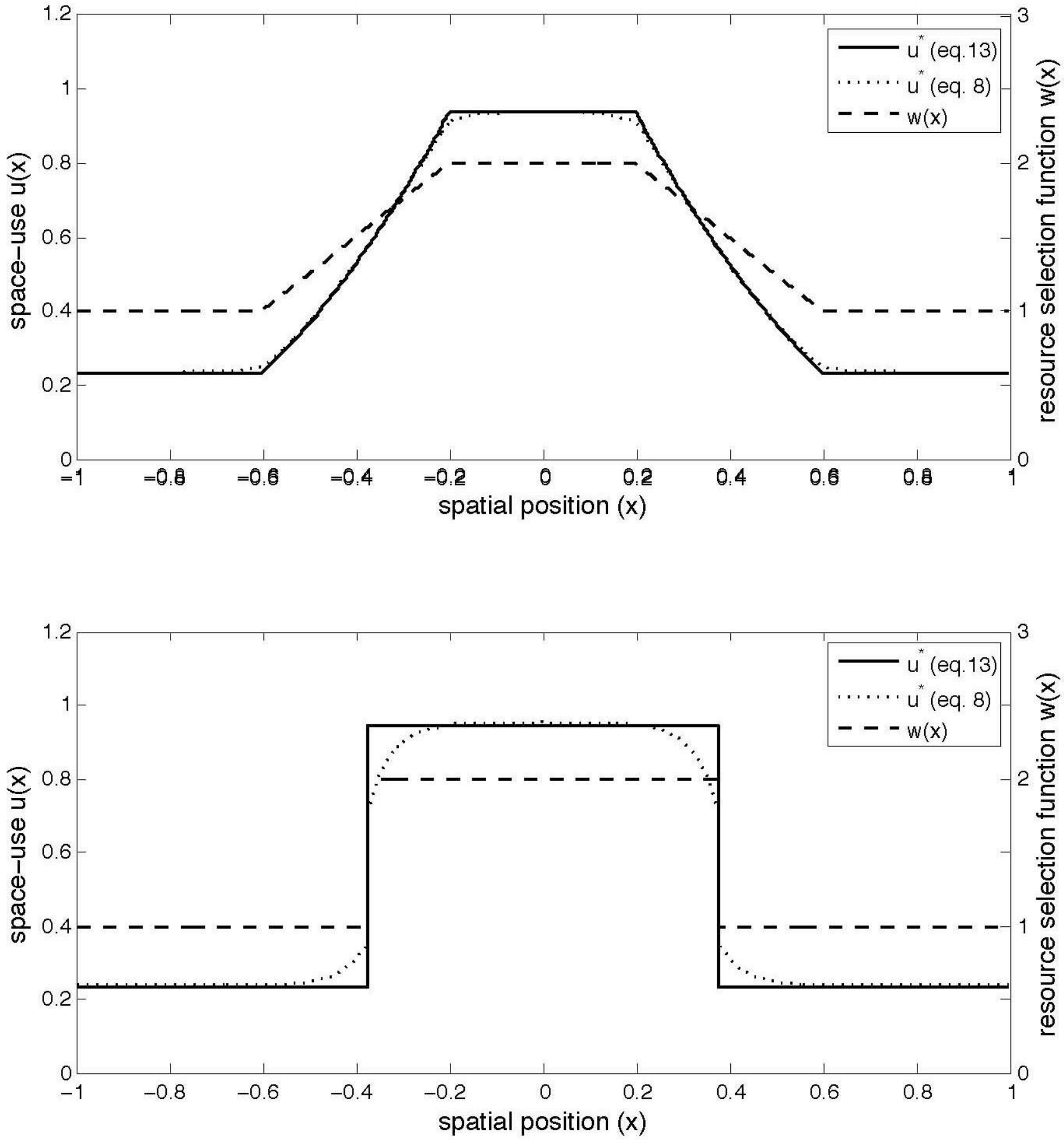

Figure 4: Patterns of space-use arising for an individual with resource selection function that is (a) piecewise continuous, and (b) discontinuous, and who moves to the left or right with equal probability with an exponential distribution of step lengths in the absence of habitat preference (i.e. $\phi_\tau(q) = (1/2l)\exp(-|q|/l)$, with $l = 0.05$ for the example shown). In both (a) and (b), the resource selection function *w(x)* (dashed line), the predicted steady-state pattern of space use $u^*(x)$ given by Eq. (13) (solid line), and $u^*(x)$ calculated by a numerical steady-state solution of Eq. (8) (dotted line) are plotted. When *w(x)* is piecewise continuous (panel a), the solution of Eq. (13) closely approximates the steady-state pattern of space-use. When *w(x)* is discontinuous (panel b), the solution of Eq. (13) fails to capture the pattern of space use in the neighborhood of the discontinuities, but still gives a reasonable match to the overall pattern of space use.

# Appendix

## Derivation of the Fokker-Planck equation for space-use

Here we derive equation Eqs. (9)-(10) from Eq. (8), for the case of general redistribution kernel $k_\tau(x,x')$[1]. First we note that in order for $u(x,t)$ to remain a probability density function, the kernel must satisfy

$$\int k_\tau(x,x')dx = 1, \qquad \text{for all } x'. \tag{A1}$$

Recalling the definition $q := x - x'$, we now change the variables used to describe the kernel from $(x,x')$ to $(q,x')$, by defining

$$\kappa_\tau(q,x') := k_\tau(x'+q,x'). \tag{A2}$$

The purpose of this is to enable us to hold $q$ constant while expanding a Taylor series in $x'$. Writing Eq. (8) in terms of this new kernel, then changing integration variable from $x'$ to $q$ gives

$$\begin{aligned} u(x,t+\tau) &= \int \kappa_\tau(x-x',x')u(x',t)dx' \\ &= \int \kappa_\tau(q,x-q)u(x-q,t)dq \\ &= \int \left[\kappa_\tau(q,x)u(x,t) - q\frac{\partial}{\partial x}[\kappa_\tau(q,x)u(x,t)] + \frac{q^2}{2!}\frac{\partial^2}{\partial x^2}[\kappa_\tau(q,x)u(x,t)]\cdots\right]dq. \end{aligned} \tag{A3}$$

The final step is achieved by considering the integrand as a function of two variables, $q$ and $(x-q)$, and Taylor expanding this function with respect to the second variable about the value $x$, while holding the first constant.

Dividing (A3) by $\tau$, making use of $\int \kappa_\tau(q,x)dq = 1$ which is a re-statement of Eq. (A1), and switching the order of differentiation and integration we get

$$\frac{u(x,t+\tau)-u(x,t)}{\tau} = -\frac{1}{\tau}\frac{\partial}{\partial x}\int q\kappa_\tau(q,x)dq \;\; u(x,t) + \frac{1}{2\tau}\frac{\partial^2}{\partial x^2}\int q^2\kappa_\tau(q,x)dq \;\; u(x,t)\cdots \tag{A4}$$

Taking the limit of small time interval $\tau$ gives

$$\begin{aligned} \frac{\partial u(x,t)}{\partial t} &= -\frac{\partial}{\partial x}\left[\lim_{\tau\to 0}\frac{1}{\tau}\left(\int_{-\infty}^{\infty} q\kappa_\tau(q,x)dq\right)\; u(x,t)\right] \\ &+ \frac{\partial^2}{\partial x^2}\left[\lim_{\tau\to 0}\frac{1}{2\tau}\left(\int_{-\infty}^{\infty} q^2\kappa_\tau(q,x)dq\right)\; u(x,t)\right] - \ldots \end{aligned} \tag{A5}$$

[1]This is essentially the derivation of a Fokker-Planck equation from a Chapman-Kolmogorov master equation via the Kramers-Moyal expansion – see Gardiner 1983.

The two integrals in the first and second terms are, respectively, the first and second moments of the kernel $\kappa_\tau$. Considering the limit of small time-steps ($\tau \to 0$), and making the conventional assumptions that the first two moments scale with order $\tau$ and that higher order terms can be discarded, we arrive at Eq (9),

$$\frac{\partial u(x,t)}{\partial t} = -\frac{\partial}{\partial x}\left[c(x)u(x,t)\right] + \frac{\partial^2}{\partial x^2}\left[d(x)u(x,t)\right], \tag{A6}$$

where

$$\begin{aligned} c(x) &= \lim_{\tau\to 0} \frac{1}{\tau}\int_{-\infty}^{\infty} q\kappa_\tau(q,x)dq, \quad \text{and} \\ d(x) &= \lim_{\tau\to 0} \frac{1}{2\tau}\int_{-\infty}^{\infty} q^2\kappa_\tau(q,x)dq. \end{aligned} \tag{A7}$$

By substituting for $q$ in the above expressions we obtain Eq. (10).

## Coefficients of the space-use equation arising from a spatially-dependent resource selection model

The redistribution kernel for the spatially-explicit resource selection model (Eq.7) is given by

$$k_\tau(x,x') = \frac{\phi_\tau(x-x')w(x)}{\int \phi_\tau(x''-x')w(x'')dx''} \tag{A8}$$

where $\phi_\tau(x-x')dx'dx$ the probability that an individual located between $x'$ and $x'+dx$ will move to a location between $x$ and $x+dx$ away from $x$, and $w(x)$ is a resource selection function. The denominator is a normalizing factor that ensures that Eq. (A1) holds.

We consider the case of continuous, sufficiently smooth positive preference function $w(x)$, and a bounded symmetric distribution of displacement distances $\phi_\tau(q)$ (Figure 3). Recalling the definition (A2), we can rewrite (A8), as

$$\kappa_\tau(q,x) = \frac{\phi_\tau(q)w(x+q)}{\int \phi_\tau(q')w(x+q')dq'} \tag{A9}$$

The $p^{th}$ moment of the distribution of displacement distances is

$$M_p(\tau) := \int q^p \phi_\tau(q)dq. \tag{A10}$$

Since $\phi_\tau$ is a probability density function $M_0(\tau)=1$, and since it is symmetric $M_1(\tau)=0$ and all higher odd moments are zero. $M_2(\tau)$ is the variance, and we make the conventional assumption that in the limit $\tau \to 0$ the higher even moments can be neglected (*i.e.* they vanish faster than linearly in $\tau$).

Under the above assumptions, Taylor expanding $w$ yields the following expression for the denominator of (A9),

$$\int \phi_\tau(q')w(x+q')dq' \quad = \quad w(x) + \frac{w_{xx}(x)}{2!}M_2(\tau) + \cdots \tag{A11}$$

where $w_{xx} = \frac{d^2w}{dx^2}$. In the limit $\tau \to 0$, only the first term of Eq. A11 remains. Using this expression for the denominator and inserting (A9) into Eqs. (A7) yields

$$\begin{aligned} c(x) &= \lim_{\tau\to 0} \frac{1}{\tau} \frac{w_x(x)M_2(\tau) + w_{xxx}(x)M_4(\tau)/3! + \cdots}{w(x) + w_{xx}(x)M_2(\tau)/2! + \cdots} \\ &= \lim_{\tau\to 0} \frac{M_2(\tau)}{\tau} \cdot \frac{w_x(x)}{w(x)} \end{aligned} \tag{A12}$$

where $w_x = \frac{dw}{dx}$, etc. Similarly,

$$\begin{aligned} d(x) &= \lim_{\tau\to 0} \frac{1}{2\tau} \frac{w(x)M_2(\tau) + w_{xx}(x)M_4(\tau)/2! \cdots}{w(x) + w_{xx}(x)M_2(\tau)/2 + \cdots} \\ &= \lim_{\tau\to 0} \frac{M_2(\tau)}{2\tau}. \end{aligned} \tag{A13}$$

This is Eq (12a,b).

As two examples, we consider two special cases of the distribution of displacement distances. Both contain a length-scale $L$ which we choose to scale as $L = \tau^{1/2}$. The distributions are (i) the fixed step length $\phi_\tau(q) = (\delta(q-L) + \delta(q+L))/2$ for which $\lim_{\tau\to 0} \frac{M_2(\tau)}{\tau} = 1$, and, (ii) the Laplace distribution (Figure 3) $\phi_\tau(q) = (1/2L)e^{-|q|/L}$, which gives $\lim_{\tau\to 0} \frac{M_2(\tau)}{\tau} = 2$. As is clear from (A12) and (A13), it is only the value of the second moment of $\phi_\tau$ that determines the magnitude of the advection and diffusion coefficients.

## Steady-state distribution of the resource selection model

The steady state distribution $u^*$ for Equation (9) can be derived by first expressing the equation in conservation law form

$$\frac{\partial u}{\partial t} + \frac{\partial}{\partial x}\left[(du)_x - cu\right] = 0. \tag{A14}$$

where the quantity in square brackets is the flux. Setting the time derivative to zero implies that the steady-state flux is a constant independent of spatial position. Since by definition the steady-state pattern of space-use $u^*$ does not change with time, this constant is zero. Thus in the steady state $(du)_x = cu$, which may be integrated to give the steady state pattern of space use

$$u^*(x) = \frac{C}{d(x)} \exp \int^x \frac{c(x')}{d(x')} dx' \tag{A15}$$

with $C$ chosen to so that $u^*(x)$ integrates to 1. Substituting (A12) and (A13) into A15 gives Eq. (13).

Note that if the individual is moving in a finite, bounded region, then zero-flux boundary conditions apply to (A6) at the edges of the domain, *i.e.*

$$\frac{\partial}{\partial x}\left[(d(x)u(x,t)\right] - c(x)u(x,t) = 0, \qquad \text{at} \quad \partial\Omega \tag{A16}$$

In this case a nonzero steady-state solution Eq. (A15) is guaranteed to exist for any continuous $w(x)$.